\documentclass[12pt]{article}
\usepackage[utf8]{inputenc}
\usepackage[english]{babel}
\usepackage{lscape}
\newcommand{\lb}[1]{\label{#1}}
\newcommand{\bi}[1]{\bibitem{#1}}
\begin{document}

\begin{center}
\begin{large}
{\bf Model independent Breit-Wigner parameters of nucleon resonances $S_{11}(1535),\ S_{11}(1650)$ and $P_{11}(1710)$}\\[0.2cm]
\end{large}
E.V.Balandina, E.M.Leikin, N.P.Yudin{\footnote{deceased}}\\[0.2cm]
Institute of Nuclear Physics, Moscow State University, Russia
\end{center}

\begin{footnotesize}
\noindent Estimates of Breit-Wigner parameters of nucleon resonances were obtained by
phenomenological analysis of $\eta$-meson photoproduction on protons  performed completely by  statistical procedures without appealing to theoretical models.\\[0.3cm]

\end{footnotesize}
We have used the results of our energy independent analysis of angular distributions of the process $\gamma p \rightarrow \eta p\,$ \cite{bal1}--\cite{bal3}. The analysis was based on expansion the differential cross sections $d\sigma(\theta)/d\Omega$ measured at definite CM energy $W$ in series in Legendre polynomials 
$P_i(\cos \theta)\,:$
$$
{{k}\over{q}}{{d\sigma(\theta)}\over{d\Omega}}=
\sum a_iP_i(\cos \theta)\,,
$$
where $k$ and $q$ are CM momenta of gamma quantum and $\eta$-meson respectively.
The use of such nonparametric statistical model provides unbiased estimates of the coefficients $a_i$ where $a_0$ is the sum of squares of the multipole amplitudes of photoproduction process. The values $a_0$ as function of CM energy $W$ and its multipole decomposition were given in  \cite{bal1}--\cite{bal3}. In present analysis we use mainly the experimental results of GRAAL collaboration  obtained in energy region $W$ from 1490 to 1724 MeV \cite{graal1}. This energy region contains the following nucleon resonsnces with the same quantum numbers of spin and isospin: $S_{11}(1535),\ S_{11}(1650)$ and $P_{11}(1710).$ 

The main task of the  analysis is to describe the energy dependence of $a_0(W)$ using statistical models based on Breit-Wigner formulae similar to \cite{chiang1}. The  fitted parameters of resonances were mass $W_R,$ full width $\Gamma_R$ and helysity amplitude $A_{1/2}.$

To keep the possibility of testing the statistical hypotheses  we  devided the whole energy region by intervals in accordence with measured in \cite{graal1}. In every interval the statistical quality of description of resonances by  parametric Breit-Wigner models with one, two and three resonances: $S_{11}(1535),\ S_{11}1535)+S_{11}(1650)$ and $S_{11}1535)+S_{11}(1650)+P_{11}(1710)$ was tested. The results of this  test based on criterion $\chi ^2/\nu$ are presented in table 1.

\begin{table}[ht]
\caption{\lb{tabl1} \small{Values of criterion $\chi^2/\nu\,,$
used in procedure of hypotheses testing }}
\begin{center}
\begin{tabular}{|l|c|c|c|c|c|c|c|}
\hline
References&\multicolumn{6}{|c|}{\cite{graal1}}&\cite{graal1,graal2}\\
\hline
\multicolumn{1}{|l|}{Energy interval }&1490 -- &1490 -- &1490 -- &1490 -- &1490 -- &1490 -- &1490 -- \\
\multicolumn{1}{|l|}{$\Delta W,$ MeV}&1584&1603& 1622& 1659&1676&1716&1724\\
\hline 
Model with  &&&&&&&\\
single resonance&1.48&1.29& 19.4&&&&\\ \hline
with two resonance &&&&1.67&1.3&8.96&\\ \hline
with three resonance&&&&&&1.36&3.97\\
\hline
\end{tabular}
\end{center}
\end{table}

\begin{table}[ht]
\caption{\lb{tabl2} \small{Values of Breit-Wigner parameters of nucleon resonances $S_{11}(1535)\,(\beta_{\eta N}=0.55,\,\beta_{\pi N}=0.35)$, $S_{11}(1650)\,(\beta_{\eta N}=0.08,\,\beta_{\pi N}=0.77)$ and $P_{11}(1710)\,(\beta_{\eta N}=0.06,\,\beta_{\pi N}=0.15).$ Here $\beta_{iN}$ are  fractions $\Gamma_{iN}/\Gamma_R$.}}
\begin{center}
\begin{small}
\begin{tabular}{|c|c|c|c|c|c|}
\hline
\multicolumn{2}{|c|}{References}&\multicolumn{3}{|c|}{\cite{graal1}}&\multicolumn{1}{|c|}{\cite{graal1,graal2}}\\
\hline
\multicolumn{2}{|c|}{$\Delta W,$ MeV}&1490-- 1603&1490 -- 1676&1490 -- 1716& 1490 -- 1724.\\
\hline 
&$W_R$, MeV&$1538.62 \pm 0.69$&$1538.06 \pm 1.12$&$1538.64 \pm 1.77$&$1539.47 \pm 4.15$\\[0.3cm]
$S_{11}(1535)$&$\Gamma_R$, MeV&$163.0 \pm 4.09$&$163.27 \pm 6.45$&$169.16 \pm 9.21$&$174.79 \pm 19.49$\\
&$A_{1/2},$ ${10^{-3}}\over {\sqrt{\rm GeV}}$&$99.23 \pm 1.01$&$101.44 \pm 2.37$&$105.66 \pm 5.44$& $112.02 \pm 18.23$\\[0.3cm]
\hline
&$W_R,$ MeV&&$1636.58 \pm 1.4$&$1640.31 \pm 4.2$&$1642.74 \pm 6.27$\\
$S_{11}(1650)$&$\Gamma_R,$ MeV&&$78.24 \pm 11.2$&$110.63 \pm 30.81$&$145.06 \pm 78.72$\\
&$A_{1/2},$ ${10^{-3}}\over {\sqrt{\rm GeV}}$&&$36.2 \pm 5.8$&$61.96 \pm 28.99$&$99.47 \pm 111.73$\\[0.3cm]
\hline
&$W_R,$ MeV&&&$1708.75\pm 1.71$&$1712.88 \pm 2.03$\\
$P_{11}(1710)$&$\Gamma_R,$ MeV&&&$45.47 \pm 15.73$&$48.74 \pm 18.8$\\
&$A_{1/2},$ ${10^{-3}}\over {\sqrt{\rm GeV}}$&&&$59.28 \pm 19.03$&$72.3 \pm 32.89$\\[0.3cm]
\hline
&$\chi ^2/\nu$&1.29&1.3&1.36&3.97\\
\hline
\end{tabular}
\end{small}
\end{center}
\end{table}

Table 2 contains the estimates of fitted parameters of nucleon resonances selected accordingly table 1 as the most reliable. Table 1 and 2 contain also the results obtained with inclusion additional experimental data of GRAAL \cite{graal2}. In all cases the fitted values  are within the limits of statistical uncertances.  Systematic errors in all cases were not taken into account. It should be noted that all   procedures used are completely statistical,  providing model independent estimates of  Breit-Wigner parameters of three nucleon resonsnces $S_{11}(1535),\ S_{11}(1650)$ and $P_{11}(1710).$ At the same time the values of Breit-Wigner parameters of nucleon resonances obtained by means of  various theoretical models with some free parameters, for example, various versions of isobar model \cite{chiang1,chiang2}, dispersion relation \cite{aznauryan}, quark models \cite{saghai} etc are not only model dependent but lead to contradictory results (for instance \cite{aznauryan}).

We hope that model independent data on nucleon resonance parameters may prove to be useful in  view of the problem of resonsnce nature \cite{krusche}.

\end{document}